\documentclass[12pt]{iopart}

\usepackage{amssymb}
\usepackage{amsthm}

 \newtheorem{theorem}{Theorem}[section]

       \newtheorem{corollary}[theorem]{Corollary}
    \newtheorem{definition}[theorem]{Definition}
       
\newcommand{\ud}{\mathrm{d}}
\begin{document}

%\title[R\'{e}nyi's Recipe and Nonextensitivity]{R\'{e}nyi's
%Recipe and Nonextensitivity: A Characterization Theorem for Tsallis
%Entropy}

\title[]{Uniqueness of Nonextensive entropy under \\ R\'{e}nyi's Recipe}

\author{Ambedkar Dukkipati\footnote{Corresponding author}, M Narasimha
Murty and Shalabh Bhatnagar}

\address{Department of Computer Science and Automation,
Indian Institute of Science, Bangalore-560012, India.}
\ead{\mailto{ambedkar@csa.iisc.ernet.in},
\mailto{mnm@csa.iisc.ernet.in}, \mailto{shalabh@csa.iisc.ernet.in}}

%----------------------------------------
\begin{abstract}
	By replacing linear
        averaging in Shannon entropy with Kolmogorov-Nagumo
        average (KN-averages) or quasilinear mean and further
        imposing the additivity   
        constraint, R\'{e}nyi proposed the first formal generalization of
        Shannon entropy. Using this recipe of R\'{e}nyi, one can prepare only
        two information measures: 
	Shannon and R\'{e}nyi entropy. Indeed, using this formalism
        R\'{e}nyi characterized these additive entropies in terms of 
        axioms of quasilinear mean. As additivity is a characteristic
        property of Shannon entropy, pseudo-additivity of the form $x \oplus_{q}
        y = x + y + (1-q)x y$ is a characteristic property of
        nonextensive (or Tsallis)
        entropy.
	One can apply R\'{e}nyi's recipe in the nonextensive case by
        replacing the linear averaging in
	Tsallis entropy with KN-averages and thereby imposing the
	constraint of 
	pseudo-additivity.
	In this paper we show that nonextensive entropy is unique
        under the R\'{e}nyi's recipe, and there by give a
        characterization.
\end{abstract}

%Uncomment for PACS numbers title message
% Keywords required only for MST, PB, PMB, PM, JOA, JOB? 
%\vspace{2pc}
%\noindent{\it Keywords}: Article preparation, IOP journals
% Uncomment for Submitted to journal title message
%\submitto{\JPA}
% Comment out if separate title page not required
\maketitle

%=========================Introduction===========================
\section{Introduction}

	In recent years, interest in generalized information measures
         has increased dramatically, after the introduction of
         {\em nonextensive entropy} in Physics
         in 1988 by
         Tsallis~\cite{Tsallis:1988:GeneralizationOfBoltzmannGibbsStatistics}.
	One can get this nonextensive entropy or Tsallis entropy by
         generalizing the information of single 
	event in the definition of Shannon entropy, by replacing
	logarithm with so
	called $q$-logarithm, 
	which is defined as 
        $\ln_{q} x = \frac{x^{1-q}-1}{1-q}$. Tsallis entropy does not
         satisfy the additivity property which is a characteristic
         property of Shannon entropy.  Instead, it satisfies
         pseudo-additivity of the form 
	$x \oplus_{q} y = x + y + (1-q)xy$ and this
         definition of entropy 
        (also known as nonextensive entropy) led
        to the field of nonextensive statistical mechanics in
        Physics. In this paper we use the term pseudo-addition to
         represent the binary operation $x \oplus_{q} y = x + y +
         (1-q)xy$ for any $q \in \mathbb{R}$ and $q > 0$.

	 Tsallis entropy is considered as a useful
         measure in describing the thermostatistical properties of a
         certain class of physical systems that entail long-range
         interactions, long-term memories and multi-fractal structures.
	 Tsallis entropy is also studied in information theory and
         Shannon-Khinchin axioms have been generalized to
         nonextensive case. While 
         canonical distributions resulting from maximization of
         Shannon entropy are exponential in nature, in the
         Tsallis case, these result in power-law distributions. To a great extent, the success of Tsallis proposal is due to
	the ubiquity of power law distributions in nature.

	Indeed, the starting point of the theory of generalized measures of
	information is due to Alfred
	R{\'{e}}nyi~\cite{Renyi:1960:SomeFundamentalQuestionsOfInformationTheory,Renyi:1961:OnMeasuresOfEntropyAndInformation}.
	By using Kolmogorov-Nagumo averages (KN-average) 
	R\'{e}nyi introduced a
	generalized information measure, known as $\alpha$-entropy or
	R\'{e}nyi entropy, the first formal well-known generalization
	of Shannon entropy.
        {\em KN-average} or quasilinear mean (we use these two
	terms interchangeably) is of
        the form
	${\langle x 
        \rangle}_{\psi} = \psi^{-1} \left (\sum_{k} p_{k}
        \psi(x_{k})\right)$,
	where $\psi$ is an arbitrary continuous
        and strictly monotone function.
	Replacing linear
        averaging in Shannon entropy with KN-averages and further
	imposing the additivity   
        constraint -- a characteristic property of underlying
        information associated with single event, which is
        logarithmic -- leads to {\em R\'{e}nyi
        entropy}. 
	Using this recipe of R\'{e}nyi, one can prepare only
        two information measures: 
	Shannon and R\'{e}nyi entropy. Using this formalism
        R\'{e}nyi characterized these additive entropies in terms of
        axioms of KN-averages.

	One can apply R\'{e}nyi's recipe in the nonextensive case by
        replacing the linear averaging in
	Tsallis entropy with KN-averages and thereby imposing the
	constraint of 
	pseudo-additivity. 
	A natural question arises: what are all the pseudo-additive
	information measures one can prepare with this recipe? We
	prove that only Tsallis entropy is possible in this case,
	which allows us to characterize
	Tsallis entropy based on axioms of KN-averages.

%	Tsallis and R{\'{e}}nyi entropy measures are two possible
%	different generalization of the Shannon entropy but are not
%	generalizations of each other.

	To understand these generalizations, the so called Hartley
        function~\cite{Hartley:1928:TransmissionOfInformation} of a
        single stochastic event plays a fundamental role. We discuss
        Hartley function in
        \S~\ref{Section:KN-avearagesAndInformationMeasures} along 
        with a brief discussion on quasilinear mean and R\'{e}nyi
        entropy. The main results of this paper, on uniqueness of Tsallis
        entropy under R\'{e}nyi's recipe and a result on
        characterization of Tsallis entropy are presented in
        \S~\ref{Section:RenyisRecipieAndTsallisEntropy} and 
        \S~\ref{Section:AcharacterizationTheoremForTsallisEntropy}
        respectively.  

%====================================================================
\section{KN-averages and Information measures}
\label{Section:KN-avearagesAndInformationMeasures}

  \subsection{Hartley Function and Shannon Entropy}

	Let $X$ be a discrete random variable (r.v) defined on some
	probability space, which takes only $n$ values, $n < \infty$.  
	We denote the set of all such random
	variables by $\mathcal{X}$. Corresponding
	to the $n$-tuple $(x_{1}, \ldots, x_{n})$ of values which $X$
	takes, probability mass function (pmf) of
	$X$ is denoted by $p = (p_{1}, \ldots p_{n})$, where $p_{k}
	\geq 0$ for $k = 1, \ldots n$ and $\sum_{k=1}^{n} p_{k}
	=1$. Expectation of r.v $X$ is denoted by $EX$ or $\langle X
	\rangle$; in this paper we use both the notations,
	interchangeably. 

	Shannon entropy, a logarithmic measure of information on $X$ denoted by $S(X)$,
	reads~\cite{Shannon:1948:MathematicalTheoryOfCommunication_BellLabs} 
	\begin{equation}
	\label{Equation:DefinitionOfShannonEntropy}
	S(X) = - \sum_{k=1}^{n} p_{k} \ln p_{k} \enspace,		
	\end{equation}
	and measures the average lack of information that is
	inherent in $p$. 

	This motivation to quantify information in terms of logarithmic
	functions is due to
	Hartley~\cite{Hartley:1928:TransmissionOfInformation}, who
	first used a logarithmic function to define uncertainty
	associated with a finite set.
	This is known as Hartley information measure.
        The Hartley information measure  of a 
        finite set $A$ with $n$ elements is defined as
        $H(A) = \log_{b} n$.
        If the base of the logarithm is $2$, then the uncertainty is
        measured in {\em bits}, and in the case of natural logarithm,
	the unit is nats. Throughout this paper we use only natural
	logarithm as a convention. 

	One can give a more general definition of Hartley information
	measure, which is a special case of Shannon entropy as
	follows. Define a function $H:
	\{x_{1}, \ldots, x_{n}  \} \rightarrow \mathbb{R}$ of the
	values taken by r.v $X \in \mathcal{X}$ with corresponding
	p.m.f $p = (p_{1}, \ldots p_{n})$ 
	as~\cite{AczelDaroczy:1975:OnMeasuresOfInformationAndTheirCharacterization} 
	\begin{equation}
	\label{Equation:HartleyFunctionForRV}
	H(x_{k}) = \ln \frac{1}{p_{k}} \enspace,\:\: \forall k = 1, \ldots n.
	\end{equation}
	$H$ is also known as entropy of a single event and plays an
	important role in all classical measures of information. It can be
	interpreted either as a measure of how unexpected the event was,
	or as measure of the information yielded by the event.
	Hartley function satisfies: (i) H is {\em
	nonnegative}: $H(x_{k})  \geq 0$ (ii) H is {\em additive}:
	$H(x_{i}x_{j}) = H(x_{i}) + H(x_{j})$ (iii) H is {\em
	normalized}:  $H(x_{k}) = 1$, whenever $p_{k} = \frac{1}{e}$
	(in the case of logarithm with 
	base $2$, the same satisfied for $p_{k} = \frac{1}{2}$). These properties
	are both necessary and
	sufficient~\cite{AczelDaroczy:1975:OnMeasuresOfInformationAndTheirCharacterization}. 

	Now, Shannon
	entropy~(\ref{Equation:DefinitionOfShannonEntropy}) can be
	written as expectation of Hartley 
	function as   
	\begin{equation}
	\label{Equation:Definition_ShannonEntropy}
	     S (X) = {\langle H \rangle} = \sum_{k=1}^{n} p_{k} H_{k} \enspace,
	\end{equation}
	where $H_{k} = H(x_{k}),\: \forall k = 1, \ldots n$, with the
	understanding that ${\langle H \rangle} = {\langle H(X)
	\rangle}$.
	
	The characteristic additive property of Shannon entropy
	\begin{equation}
	\label{Equation:AdditivityOfShannonEntropy}
	     S(X \times Y) = S(X) + S(Y) \enspace,
	\end{equation}
	for two independent random variables $X$ and
	$Y$ now follows as a consequence of the additivity property of
	Hartley function. 

	There are two postulates involved in defining Shannon entropy
	as expectation of Hartley function. One is the additivity of
	information which is the characteristic property of Hartley
	function, and the other is
	that if different amounts of information occur with different
	probabilities, the total information will be the
	average of the individual informations weighted by the
	probabilities of their occurrences. 

	The basic idea behind R\'{e}nyi's generalization is any
        putative candidate for an entropy should be a mean and there
        by use a  well known
        idea in mathematics   
        that the linear mean, though most widely used, is not the only
        possible way of averaging, however, one can define the mean with
        respect to an arbitrary
        function. Here we briefly discuss
        generalized averages and its properties which are essential for
        the results we present in this paper.

  %-----------------------------------------------------------------
  \subsection{Kolmogorov-Nagumo Averages or Quasilinear Mean}

        In the general theory of means, quasilinear mean of a random variable
        $X$ is defined as{\footnote{Kolmogorov~\cite{Kolmogorov:1930:SurLaNotionDeLaMoyenne} and Nagumo~\cite{Nagumo:1930:UberEineKlasseVonMittlewerte}
        first characterized the quasilinear mean ${\langle x
        \rangle}_{\psi}$ for a vector $(x_{1}, \ldots,
        x_{n})$ as ${\langle x \rangle}_{\psi} =
        \psi^{-1}\left(\sum_{k=1}^{n} \frac{1}{n} \psi(x_{k})\right)$
        where $\psi$ is a continuous and strictly monotone
        function. De Finetti~\cite{DeFinetti:1931:SulConcettoDiMedia}
        extended their result to the case of simple (finite)
        probability distributions. The version of the quasilinear mean
        representation theorem referred to in
        \S~\ref{Section:AcharacterizationTheoremForTsallisEntropy} is
        due to Hardy, Littlewood and
        P{\'{o}}lya~\cite{HardyLittlewoodPolya:1934:Inequalities}, which
        followed closely the approach of de
        Finetti. Acz{\'{e}}l~\cite{Aczel:1948:OnMeanValues} proved a
        characterization of the quasilinear mean using functional
        equations.
        Ben-Tal~\cite{Ben-Tal:1977:OnGeneralizedMeansAndGeneralizedConvexFucntions}
        showed that quasilinear means are ordinary arithmetic means
        under suitably defined addition and scalar multiplication
        operations.
        Norris~\cite{Norris:1976:GeneralMeansAndStatisticalTheory} did
        a survey of quasilinear means and its more restrictive forms in
        Statistics. More recent survey of generalized means can be
        found
        in~\cite{OstasiewiczOstasiewicz:2000:MeansAndTheirAppliacations}.
	Applications of quasilinear means can be found in economics
        (for example, 
        \cite{EpsteinZin:1989:SubstitutionRisk_SecondaryRef}) and 
        decision theory (for example,
        \cite{KrepsPorteus:1978:TemporalResolution_SecondaryRef}).
        Recently Czachor and
        Naudts~\cite{CzachorNaudts:2002:ThermostatisticsBasedOnKolmogorov-NagumoAverages}
        studied generalized thermostatistics based on quasilinear means.}%ENDfootnote
        \begin{equation}
        \label{Equation:Definition_KNaverages}
         E_{\psi}X = {\langle X \rangle}_{\psi} = \psi^{-1} \left( \sum_{k=1}^{n}
        p_{k} \psi\left(x_{k} \right)    \right) \enspace,
        \end{equation}
        where $\psi$ is continuous and strictly monotonic (increasing
        or decreasing) in which
        case it has an inverse $\psi^{-1}$ which satisfies the same
        conditions. In the context of generalized means, $\psi$ is
        referred to as Kolmogorov-Nagumo 
        function or KN-function.
        If, in particular, $\psi$ is linear, then 
        (\ref{Equation:Definition_KNaverages}) reduces to the
        expression of linear averaging,
        $EX = {\langle X \rangle} = \sum_{k=1}^{n} p_{k} x_{k}$.

        The following theorem qualifies quasilinear means.
	%THEOREM:KN-average as a Mean----
        \begin{theorem}
        \label{Theorem:KN:KNaverageAsMean}
                If  $\psi$ is continuous and strictly monotone in
                        $a \leq x \leq b$, $a \leq x_{k} \leq b,\:\:\:
        k = 1, \ldots n$, $p_{k} > 0 $ and $\sum_{k=1}^{n} p_{k} =1 $,
        then
                $\exists$ unique $x_{0} \in (a,b)$ such that
                \begin{displaymath}
                 \psi(x_{0}) = \sum_{k=1}^{n} p_{k} \psi(x_{k})
                \end{displaymath}
                and $x_{0}$ is greater than some and less than
                others of the $x_{k}$ unless all $x_{k}$ are zero.
        \end{theorem}

        Thus, the mean ${\langle \, . \,\rangle}_{\psi}$ is determined when the
        function $\psi$ is given. We may ask whether the converse is
        true: if ${\langle X \rangle}_{\psi_{1}} ={\langle
        X \rangle}_{\psi_{2}} $ for all $X \in \mathcal{X}$, is
        $\psi_{1}$ 
        necessarily the same function as $\psi_{2}$?  
        First we give the following definition.
        %DEFINITION:Equivalent Mean-----
        \begin{definition}
        \label{Definition:KNequivalentFunctions}
        Continuous and strictly monotone functions $\psi_{1}$ and $\psi_{2}$ are
        said to be {\em KN-equivalent} if ${\langle X \rangle}_{\psi_{1}} =
        {\langle X \rangle}_{\psi_{2}}$ for all $X \in \mathcal{X}$.
        \end{definition}
        Note that when we compare two means, it is to be understood
        that the underlying probabilites are same. The following
        theorem characterizes KN-equivalent functions.
        %THEOREM:Condition for KN-equivalent Functions
        \begin{theorem}
        \label{Theorem:ConditionForKNequivalentFuntions}
        In order that two continuous and strictly monotone functions
        $\psi_{1}$ and $\psi_{2}$ are KN-equivalent, it is necessary and sufficient
        that
        \begin{displaymath}
                \psi_{1} = \alpha \psi_{2} + \beta \enspace,
        \end{displaymath}
        where $\alpha$ and $\beta$ are constants and $\alpha \neq 0$.
        \end{theorem}
        
        \begin{corollary}
        Let $\psi$ be a KN-function then ${\langle X \rangle}_{\psi} =
        {\langle X \rangle}_{-\psi}$ .
        \end{corollary}
        Hence, when ever required, without loss of generality, one
        can assume that $\psi$ is an increasing function. 
        The following theorem characterizes additivity of quasilinear means.
        \begin{theorem}
        \label{Theorem:AdditivityOfKNaverages}
        Let $\psi$ be a KN-function and $c$ be a real constant then
        ${\langle X + c\rangle}_{\psi} = {\langle X \rangle}_{\psi} +
        c$ i.e.,
        \begin{displaymath}
         \psi^{-1} \left( \sum_{k=1}^{n}
        p_{k} \psi\left(x_{k} + c \right) \right) = \psi^{-1} \left( \sum_{k=1}^{n}
        p_{k} \psi\left(x_{k} \right) \right) + c
        \end{displaymath}
        if and only if $\psi$ is either linear or exponential.
        \end{theorem}
	Proof of
        Theorems~\ref{Theorem:KN:KNaverageAsMean},
        \ref{Theorem:ConditionForKNequivalentFuntions} and
        \ref{Theorem:AdditivityOfKNaverages}  
        can be found in the book on inequalities by Hardy, Littlewood,
        P{\'{o}}lya~\cite{HardyLittlewoodPolya:1934:Inequalities}.  

  %-----------------------------------------------------        
  \subsection{R\'{e}nyi Entropy}

        In the definition of Shannon entropy
        (\ref{Equation:Definition_ShannonEntropy}), if the standard
        mean 
        of Hartley function $H$
        is replaced with the quasilinear
        mean~(\ref{Equation:Definition_KNaverages}), one can obtain a
        generalized measure of information of r.v $X$ with respect to
        a KN-function $\psi$ as
        \begin{equation}
	\label{Equation:QuasilinearEntropy}
        S_{\psi}(X) = \psi^{-1} \left(\sum_{k=1}^{n} p_{k} \psi \left(
        \ln \frac{1}{p_{k}} \right) \right) = \psi^{-1}
        \left(\sum_{k=1}^{n} p_{k} \psi \left( 
        H_{k} \right) \right) \enspace,
        \end{equation}
        where $\psi$ is a KN-function. We refer to
        (\ref{Equation:QuasilinearEntropy}) as quasilinear entropy
        with respect to the KN-function $\psi$.
        If we impose the constraint of additivity on $S_{\psi}$, then
        $\psi$  should
        satisfy~\cite{Renyi:1960:SomeFundamentalQuestionsOfInformationTheory}  
        \begin{equation}
        \label{Equation:AdditivityEquationForKNaverages}
        {\langle X + c \rangle}_{\psi} = {\langle X \rangle}_{\psi} +
        c \enspace, 
        \end{equation}
        for any random variable $X \in \mathcal{X}$ and a constant
        $c$. 

        R\'{e}nyi employed this formalism to define a
        one-parameter family 
        of measures of information ($\alpha$-entropies) as follows:
	%Equation: Definition of Renyi entropy
        \begin{equation}
	\label{Equation:Definition_RenyiEntropy}
        S_{\alpha}(X) = \frac{1}{1-\alpha} \ln \left(\sum_{k=1}^{n}
        p_{k}^{\alpha} \right) \enspace,
        \end{equation}
        where the KN-function $\psi$ is chosen in 
        (\ref{Equation:QuasilinearEntropy}) as   
        $\psi(x) = e^{(1-\alpha)x}$ whose choice is motivated by 
        Theorem~\ref{Theorem:AdditivityOfKNaverages}. If we choose
        $\psi$ as a 
        linear function in quasilinear
        entropy~(\ref{Equation:QuasilinearEntropy}), what we get is
        Shannon entropy.  
        R\'{e}nyi entropy is a
        one-parameter generalization of Shannon entropy in the sense
        that the limit $\alpha \rightarrow 1$ in
        (\ref{Equation:Definition_RenyiEntropy}) retrieves Shannon
        entropy. 

        %applications
        Despite its formal origin R\'{e}nyi entropy proved important
        in a variety of practical applications in coding
        theory~\cite{AczelDaroczy:1975:OnMeasuresOfInformationAndTheirCharacterization},
        statistical
        inference~\cite{ArimitsuArimitsu:2000:TsallisStatisticsAndTurbulence_SecondaryRef,ArimitsuArimitsu:2001:AnalysisOfTurbulence_SecondaryRef}, quantum 
        mechanics~\cite{MaassenUffink:1988:GeneralizedEntropicUncertaintyRelations},
        chaotic dynamics
        systems~\cite{HalseyJensenKadanoffProcacciaShraiman:1986:FractalMeasuresAndThierSingularities}.
        Thermodynamic properties of systems with multi-fractal 
        structures have been studied by extending the notion of
        Gibbs-Shannon entropy into a more general framework - R\'{e}nyi
        entropy~\cite{JizbaArimitsu:2004:ObservabilityOfRenyiEntropy}.

%=============================================================
\section{R\'{e}nyi's Recipe and Tsallis Entropy}
\label{Section:RenyisRecipieAndTsallisEntropy}

  %--------------------------------------------------
  \subsection{Tsallis Entropy}

	Due to an increasing interest in long-range correlated systems
	and non-equilibrium phenomena there has recently been much
	focus on the Tsallis (or nonextensive)
	entropy. Although, first introduced by Havrda and Charvat
	\cite{HavrdaCharvat:1967:QuantificationMethodOfClassificationProcess}
	in the context of cybernetics theory 
        and later studied by
	Dar{\'{o}}czy~\cite{Daroczy:1970:GeneralizedInformationFunctions},
	it was 
	Tsallis~\cite{Tsallis:1988:GeneralizationOfBoltzmannGibbsStatistics}
	who exploited its nonextensive features and placed it in a
	physical setting. Hence it is also known as
	Harvda-Charvat-Dar\'{o}czy-Tsallis entropy. Throughout this
	paper we refer to this as Tsallis or nonextensive
	entropy. Tsallis entropy of a r.v $X \in \mathcal{X}$ with p.m.f
	$p=(p_{1}, \ldots p_{n})$ is defined as
	\begin{equation}
	\label{Equation:Definition_TsallisEntropy}
	  S_{q}(X) = \frac{1 - \sum_{k=1}^{n} p_{k}^{q}}{q-1} \enspace,
	\end{equation}
	where $q >0$ is called the nonextensive index.
	%($q$ is positive in
	%order to ensure the concavity of $S_{q}$).
	Tsallis entropy too, like R\'{e}nyi entropy, is a
	one-parameter generalization of 
	Shannon entropy in the sense that $q \rightarrow 1$ in
	(\ref{Equation:Definition_TsallisEntropy}) retrieves Shannon
	entropy. Tsallis entropy is
	concave for all $q > 0$, but R\'{e}nyi entropy is concave only
	for $0 < \alpha < 1 $.  The index $q$ characterizes the
	degree of 
	nonextensivity reflected in the pseudo-additivity property
	\begin{equation}
	\label{Equation:PseudoAdditivityOfTsallisEntropy}
	S_{q}(X \times Y) = S_{q}(X) \oplus_{q} S_{q}(Y) = S_{q}(X) + S_{q}(Y) +
	(1-q) S_{q}(X) S_{q}(Y) \enspace,
	\end{equation}
	where $X,Y \in \mathcal{X}$ are two independent random variables.

  %----------------------------------------------------------
  \subsection{Nongeneralizability of Tsallis Entropy}

	Though the derivation of Tsallis entropy, when it was proposed
	in 1988~\cite{Tsallis:1988:GeneralizationOfBoltzmannGibbsStatistics} is slightly different, one can understand this
	generalization using $q$-logarithm
	function (see~(\ref{Equation:Definition_q-Logorithm})), where
	one would first generalize logarithm in the 
	Hartley information with $q$-logarithm and define $q$-Hartley
	function $\widetilde{H}: \{x_{1}, \ldots, x_{n}\} \rightarrow
	\mathbb{R}$ of r.v $X$ as
	~\cite{Tsallis:1999:NonextensiveStatisticalMechanics} 
	\begin{equation}
	\label{Equation:Definition_q-HartleyInformationMeasure}
	\widetilde{H}_{k}=\widetilde{H}(x_{k}) = \ln_{q}
	\frac{1}{p_{k}}\enspace, \quad k=1,\ldots n \enspace.
	\end{equation}
	The $q$-logarithm
	in~(\ref{Equation:Definition_q-HartleyInformationMeasure}) is
	defined as 
	\begin{equation}
	\label{Equation:Definition_q-Logorithm}
	\ln_{q}(x) = \frac{x^{1-q}-1}{1-q} \enspace,
	\end{equation}
	which satisfies pseudo-additivity of the form
	$\ln_{q}(xy)=\ln_{q}x \oplus_{q}
	\ln_{q}y$ and in the limit $q \to 1$, we have $\ln_{q} x \to \ln x$.
	Now Tsallis entropy
	(\ref{Equation:Definition_TsallisEntropy}) 
	can be defined as the expectation of $q$-Hartley function $\widetilde{H}$
	as 
	\begin{equation}
	\label{Equation:Definition_TsallisEntropy_2}
	S_{q}(X) = {\left\langle \widetilde{H} \right\rangle} \enspace.
	\end{equation}
	Note that the characteristic pseudo-additivity property of Tsallis
	entropy~(\ref{Equation:PseudoAdditivityOfTsallisEntropy}) 
	is a consequence of additivity property of Hartley
	function. 

	Before we present the main results of this paper, we briefly
	discuss the context of quasilinear means where there is a
	relation between Tsallis and R\'{e}nyi entropy.
	The $q$-Hartley function can be written as
	\begin{displaymath}
	\widetilde{H}_{k} = \ln_{q} \frac{1}{p_{k}} = \phi_{q}(H_{k})\enspace,
	\end{displaymath}
	where
	 \begin{equation}
        \label{Equation:KN:ModfiedKNfunction}
        \phi_{q}(x) = \frac{e^{(1-q)x} -1}{1 - q} =
	\ln_{q}(e^{x}) \enspace. 
        \end{equation}
	Note that $\phi_{q}$ is KN-equivalent to $e^{(1-q)x}$
	(by Theorem~\ref{Theorem:ConditionForKNequivalentFuntions}), the 
	KN-function used in R\'{e}nyi entropy. Hence 
	Tsallis entropy is related to R\'{e}nyi entropies as 
	\begin{equation}
        \label{Equation:RelationBetweenTsallisAndRenyi_ViaKN}
	S_{q}^{\mbox{T}} = \phi_{q}(S_{q}^{\mbox{R}}) \enspace,
	\end{equation}
	where $S_{q}^{\mbox{T}}$ and $S_{q}^{\mbox{R}}$ denote the
	Tsallis and R\'{e}nyi entropy respectively with a real number
	$q$ as a parameter.
	Hence, Tsallis entropy and R\'{e}nyi entropy are monotonic
	functions of each other and, as a result, both must be
	maximized by the same probability distribution. 

	Now a natural question that arises is
	whether one could generalize Tsallis
	entropy using R\'{e}nyi's recipe i.e., by replacing linear average in
	(\ref{Equation:Definition_TsallisEntropy_2}) by KN-averages
	and impose the 
	condition of pseudo-additivity. It is equivalent to determining
	the KN-function $\psi$ for which so called $q$-quasilinear
	entropy defined as 
	\begin{equation}
	\label{Equation:Definition_q-QuasilinearEntropy}
	\widetilde{S}_{\psi} (X) = {\left\langle \widetilde{H}
	\right\rangle}_{\psi} = \psi^{-1}
	\left[ \sum_{k=1}^{n} p_{k} \psi \left( \widetilde{H}_{k}
	\right) \right] \enspace,
	\end{equation}
	where $\widetilde{H}_{k} = \widetilde{H}(x_{k})\: \forall k = 
	1, \ldots n$, satisfies the pseudo-additive property.

	First, we present the following result which characterizes the
	pseudo-additivity of quasilinear means.
	%THEOREM:Nonextensive Additivity of Two Random Variables
        \begin{theorem}
        \label{Theorem:NonextensiveAditivityOfTwoRandomVariables}
	Let $X,Y \in \mathcal{X}$ be two independent random
        variables. Let $\psi$ be any KN-function. Then 
	\begin{equation}
	\label{Equation:NonextensiveAdditivityOfKN-averages_Condition_Form1}
	{\langle X \oplus_{q} Y \rangle}_{\psi} = {\langle X \rangle}_{\psi} \oplus_{q}{\langle Y \rangle}_{\psi}
	\end{equation}
	if and only if $\psi$ is linear.
        \end{theorem}
	%PROOF....
        \proof
	Let $p$ and $r$ be the p.m.fs of random variables $X, Y \in
	\mathcal{X}$ respectively.
	The proof of
	sufficiency is simple which follows from
	\begin{displaymath}
	{\langle X \oplus_{q} Y \rangle}_{\psi} = {\langle X
	\oplus_{q} Y \rangle} = \sum_{i=1}^{n} \sum_{j=1}^{n}
	p_{i}r_{j} (x_{i} \oplus_{q} y_{j}) \enspace,
	\end{displaymath}
	and by the definition of $\oplus_{q}$, we have
	{\setlength\arraycolsep{0pt}
        \begin{eqnarray}
	{\langle X \oplus_{q} Y \rangle} &=& \sum_{i=1}^{n} \sum_{j=1}^{n}
	p_{i}r_{j} (x_{i} + y_{j} + (1-q) x_{i} y_{j}) \nonumber\\
	& = & \sum_{i=1}^{n} p_{i} x_{i} + \sum_{j=1}^{n} r_{j} y_{j}
        + (1-q) \sum_{i=1}^{n} p_{i} x_{i} \sum_{j=1}^{n} r_{j} y_{j}\enspace.
	\nonumber 
        \end{eqnarray}}

	To prove the converse, we need to determine all forms of $\psi$ which
	satisfy
        \begin{equation}
        \label{Equation:NonextensiveAdditivityOfKN-averages_Condition_Form2}
        \psi^{-1} \left(\sum_{i=1}^{n} \sum_{j=1}^{n} p_{i}r_{j} 
        \psi \left( x_{i} \oplus_{q} y_{j}
        \right)  \right)  
         = \psi^{-1} \left(\sum_{i=1}^{n} p_{i} \psi \left( x_{i}
        \right)  \right) \oplus_{q} \psi^{-1} \left(\sum_{j=1}^{n}
        r_{j} \psi \left( y_{j} \right)  \right) \enspace.
        \end{equation}

        Since~(\ref{Equation:NonextensiveAdditivityOfKN-averages_Condition_Form2})
        must hold for arbitrary p.m.fs $p$,$r$ and for arbitrary
        numbers  
        $\{x_{1}, \ldots, x_{n}\}$ and $\{y_{1}, \ldots, y_{n}\}$, one
        can choose $y_{j} = c$ independently of $j$.  Then 
        (\ref{Equation:NonextensiveAdditivityOfKN-averages_Condition_Form2})
        yields  
        \begin{equation}
        \label{Equation:NonextensiveAdditivityOfKN-averages_Condition_Form3}
        \psi^{-1} \left(\sum_{i=1}^{n} p_{k}
        \psi \left( x_{i} \oplus_{q} c \right)  \right) =
        \psi^{-1} \left(\sum_{i=1}^{n} p_{k} \psi \left(
        x_{i} \right) \right) \oplus_{q} c \enspace.
        \end{equation}
	That is, $\psi$ should satisfy
        \begin{equation}
        \label{Equation:NonextensiveAdditivityOfKN-averages_Condition_Form4}
        {\langle X \oplus_{q} c \rangle}_{\psi} = {\langle X
        \rangle}_{\psi} \oplus_{q} c \enspace,
        \end{equation}
        for any $X \in \mathcal{X}$ and any constant $c$. This can be
        rearranged as
        \begin{displaymath}
        {\langle (1 + (1-q) c) X + c \rangle}_{\psi} =
          (1 + (1-q) c) {\langle X \rangle}_{\psi} + c 
        \end{displaymath}
	by using the definition of $\oplus_{q}$.
        Since  $q$ is independent of other quantities, $\psi$ should
        satisfy an equation of the form 
        \begin{equation}
        \label{Equation:NonextensiveAdditivityOfKN-averages_Condition_Form5}
        {\langle dX + c \rangle}_{\psi} = d {\langle X \rangle}_{\psi}
        + c \enspace,
        \end{equation}
        where $d \neq 0$ (by writing $d =(1+(1-q)c)$).
        Finally $\psi$ must satisfy
        \begin{equation}
        \label{Equation:NonextensiveAdditivityOfKN-averages_Condition_Sub1}
        {\langle X + c \rangle}_{\psi} = {\langle X \rangle}_{\psi} + c
        \end{equation}
        and 
        \begin{equation}
        \label{Equation:NonextensiveAdditivityOfKN-averages_Condition_Sub2}
        {\langle dX \rangle}_{\psi} = d {\langle X \rangle}_{\psi} \enspace,
        \end{equation}
        for any $X \in \mathcal{X}$ and any constants $d$, $c$. 
        From Theorem~\ref{Theorem:AdditivityOfKNaverages}, the condition 
        (\ref{Equation:NonextensiveAdditivityOfKN-averages_Condition_Sub1}) 
        is satisfied only when $\psi$ is linear or exponential.

	To complete the theorem we have to show that
        KN-averages do not satisfy condition
        (\ref{Equation:NonextensiveAdditivityOfKN-averages_Condition_Sub2})
        when $\psi$ is exponential.
	For a particular choice of
        $\psi(x) = e^{(1- \alpha)x}$, assume that
        \begin{equation}
	\label{Equation:ToGetTheContradiction_ForTheTheorem}
        {\langle d X \rangle}_{\psi} = d {\langle X
        \rangle}_{\psi} \enspace,
        \end{equation}
        where
        \begin{displaymath}
        {\langle d X \rangle}_{\psi_{1}} = \frac{1}{1-\alpha} \ln
        \left( \sum_{k=1}^{n} p_{k} e^{(1-\alpha) d x_{k}} \right) \enspace,
        \end{displaymath}
	and
        \begin{displaymath}
        d {\langle X \rangle}_{\psi_{1}} = \frac{d}{1-\alpha} \ln
        \left( \sum_{k=1}^{n} p_{k} e^{(1-\alpha) x_{k}} \right)  \enspace.
        \end{displaymath}
        Now define a KN-function $\psi'$  as $\psi'(x) = e^{(1-
        \alpha)dx}$, for which 
        \begin{displaymath}
        {\langle X \rangle}_{\psi'} = \frac{1}{d(1-\alpha)} \ln 
        \left( \sum_{k=1}^{n} p_{k} e^{(1-\alpha) d x_{k}} \right) \enspace.
        \end{displaymath}
	Condition
        (\ref{Equation:ToGetTheContradiction_ForTheTheorem}) implies
       	\begin{displaymath}
        {\langle X \rangle}_{\psi} = {\langle X \rangle}_{\psi'} \enspace,
	\end{displaymath}
	and by 
        Theorem~\ref{Theorem:ConditionForKNequivalentFuntions},
        $\psi$ and $\psi'$ are 
        KN-equivalent which gives a contradiction.

        \endproof
	%ENDPROOF.....

	One can observe that the above proof avoids solving 
	functional equations as in the case of
	Theorem~\ref{Theorem:AdditivityOfKNaverages} (see
	\cite{AczelDaroczy:1975:OnMeasuresOfInformationAndTheirCharacterization}).
	Instead it makes
	use of basic results of KN-averages. 
	The following corollary is the immediate consequence of 
	Theorem~\ref{Theorem:NonextensiveAditivityOfTwoRandomVariables}. 
	%Theorem: Nongeneralizability of Tsallis Entropy------
        \begin{corollary}
        \label{Corollary:NongenralizabilityOfTsallisEntropy}
	$q$-quasilinear entropy $\widetilde{S}_{\psi}$ (defined as
	in~(\ref{Equation:Definition_q-QuasilinearEntropy})) with respect to
	a KN-function $\psi$ satisfies pseudo-additivity if 
        and only if $\widetilde{S}_{\psi}$ is Tsallis entropy.
        \end{corollary}
        \proof
	Let $X,Y \in \mathcal{X}$ be two independent random variables
	and let
	$p,r$ be their corresponding pmfs. 
        By the pseudo-additivity constraint, $\psi$ should satisfy
        \begin{equation}
        \label{Equation:KNtsallis_PseudoAdditivity_Condition_Form1}
        \widetilde{S}_{\psi}(X \times Y) = \widetilde{S}_{\psi}(X) \oplus_{q}
        \widetilde{S}_{\psi}(Y) 
        \end{equation}
        From the property of $q$-logarithm that $\ln_{q} x y = \ln_{q}x
        \oplus_{q} \ln_{q}y$, we need
        {\setlength\arraycolsep{0pt}
        \begin{eqnarray}
        \label{Equation:KNtsallis_PseudoAdditivity_Condition_Form2}
        \psi^{-1}  && \left(\sum_{i=1}^{n} \sum_{j=1}^{n} p_{i}r_{j} \psi
        \left( \ln_{q} \frac{1}{p_{i}r_{j}}  \right)  \right)  \nonumber\\
        && = \psi^{-1} \left(\sum_{i=1}^{n} p_{i} \psi \left( \ln_{q}
        \frac{1}{p_{i}}  \right)  \right) \oplus_{q}
        \psi^{-1} \left(\sum_{j=1}^{n} r_{j} \psi \left( \ln_{q}
        \frac{1}{r_{j}}  \right)  \right) \enspace.
        \end{eqnarray}
        Equivalently, we need
        {\setlength\arraycolsep{0pt}
        \begin{eqnarray}
        \psi^{-1} && \left(\sum_{i=1}^{n} \sum_{j=1}^{n} p_{i}r_{j} 
        \psi \left( \widetilde{H}_{i}^{p} \oplus_{q} \widetilde{H}_{j}^{r}
        \right)  \right)   \nonumber \\
         && = \psi^{-1} \left(\sum_{i=1}^{n} p_{i} \psi \left(
        \widetilde{H}_{i}^{p}   \right)  \right) \oplus_{q} 
        \psi^{-1} \left(\sum_{j=1}^{n} r_{j} \psi
        \left(\widetilde{H}_{j}^{r} \right)  \right) \enspace, \nonumber
        \end{eqnarray}
        where $\widetilde{H}^{p}$ and $\widetilde{H}^{r}$ represent
        the $q$-Hartley functions corresponding to probability distributions $p$
        and $r$ respectively.
	That is, $\psi$ should satisfy
	\begin{displaymath}
	{\langle \widetilde{H}^{p} \oplus_{q}  \widetilde{H}^{r}
	\rangle}_{\psi}  =  {\langle \widetilde{H}^{p} \rangle}_{\psi}
	\oplus_{q}  {\langle \widetilde{H}^{r} \rangle}_{\psi} \enspace.
	\end{displaymath}
	Also from 
	Theorem~\ref{Theorem:NonextensiveAditivityOfTwoRandomVariables},
	$\psi$ is linear and hence $\widetilde{S}_{\psi}$ is Tsallis.
        \endproof
	Corollary~\ref{Corollary:NongenralizabilityOfTsallisEntropy}
	shows that using the R\'{e}nyi's recipe in the nonextensive
	case one can prepare only Tsallis entropy, while in the
	classical there are two possibilities.

%=============================================================
\section{A Characterization Theorem for Tsallis Entropy}
\label{Section:AcharacterizationTheoremForTsallisEntropy}

	The importance of R\'{e}nyi's formalism to generalize Shannon
	entropy is a characterization of Shannon entropy in terms of
	axiom of quasilinear
	means~\cite{Renyi:1960:SomeFundamentalQuestionsOfInformationTheory}.
	By the result,
	Theorem~\ref{Theorem:NonextensiveAditivityOfTwoRandomVariables},
	that we presented in this paper, one can give a
	characterization of 
	Tsallis entropy in terms of axioms of quasilinear means. For such a
	characterization one would assume that entropy is the expectation
	of a function of underlying r.v. In the classical case, the
	function is Hartley function, while in the nonextensive case
	it is $q$-Hartlay function.

	Since characterization of quasilinear means is given in terms of
	cumulative distribution of a random variable, we use the 
	following definitions and notation.
	
	Let $F:{\mathbb{R}} \rightarrow
        {\mathbb{R}}$ denote the cumulative distribution function of
        random variable $X \in \mathcal{X}$. Corresponding to a
        KN-function $\psi: {\mathbb{R}} \rightarrow {\mathbb{R}}$,
        generalized mean of $F$ (or $X$) can be written as
        \begin{equation}
        \label{Equation:KN-averagesInTermsOfCumulativeDistribution}
          E_{\psi}(F)= E_{\psi}(X) = {\langle X \rangle}_{\psi} =
        \psi^{-1}\left(\int \psi \, \ud 
        F \right) \enspace,
        \end{equation}
	which is continuous analogue to
        (\ref{Equation:Definition_KNaverages}) and it is axiomized by
        Kolmogorov, Nagumo and De Finetti (see 
        \cite[Theorem 215]{HardyLittlewoodPolya:1934:Inequalities}) as
        follows.

        %Theorem: Axioms of Kolmogorov Nagumo Averages
        \begin{theorem}
        \label{Theorem:AxiomsForKN-averages}
        Let $\mathcal{F}_{I}$ be the set of all cumulative
        distribution functions defined on some interval $I$ of the
        real line ${\mathbb{R}}$. A functional $\kappa:
        {\mathcal{F}}_{I} \rightarrow {\mathbb{R}}$ satisfies the
        following axioms:
        \begin{description}
          \item[axiom 1:] $\kappa(\delta_{x}) = x$, where $\delta_{x} \in 
        {\mathcal{F}}_{I}$ denotes the step function at
        $x$ (\textit{Consistency with certainty}) ,

          \item[axiom 2:] $F,G \in
          {\mathcal{F}}_{I}$, if $F \leq G $ then $\kappa(F) \leq
          \kappa(G)$; the equality holds if and only if $F = G$
          (\textit{Monotonicity}) and,

%         \item[axiom 2:] (\textit{Substitution}) $F,G \in
%         {\mathcal{F}}_{I}$, if $E(F) = E(G)$ then
%         $\forall \beta \in (0,1) \:\: \exist \gamma \in (0,1)$ such
%         that $ E(\beta F + (1-\beta)H) = E( \gamma
%         G + (1-\gamma)H)$, for any $H \in {\mathcal{F}}_{I}$ 

          \item[axiom 3:] $F,G \in
          {\mathcal{F}}_{I}$, if $\kappa(F) = \kappa(G)$ then
          $ \kappa(\beta F + (1-\beta)H) = \kappa( \beta
          G + (1-\beta)H)$, for any $H \in {\mathcal{F}}_{I}$
          (\textit{Quasilinearity})  

        \end{description}
        if and only if
        there is a continuous strictly monotone function $\psi$ such
        that
        \begin{displaymath}
        \kappa(F) = 
        \psi^{-1}\left(\int \psi \, \ud F \right) \enspace.
        \end{displaymath}
        \end{theorem}
        
        The modified axioms for quasilinear mean can be found in
        \cite{Chew:1983:AgeneralizationOfTheQuasilinearMean,Fishburn:1986:ImplicitMeanValues,OstasiewiczOstasiewicz:2000:MeansAndTheirAppliacations}).
        Now we give our characterization theorem for Tsallis entropy
        that is similar to the
        characterization of Shannon entropy given by
        R\'{e}nyi~\cite{Renyi:1960:SomeFundamentalQuestionsOfInformationTheory}.  
        \begin{theorem}
        \label{Theorem:CharacterizationOfTsallisEntropy}
	Let $X \in \mathcal{X}$ be a random variable. An information measure 
        defined as a (generalized) mean $\kappa$ of $q$-Hartley function of
        $X$ is Tsallis entropy if and only if
        \begin{enumerate}
          \item $\kappa$ satisfies axioms of quasilinear means given in 
          Theorem~\ref{Theorem:AxiomsForKN-averages} and, 

          \item If $X,Y \in \mathcal{X}$ are two random variables which 
          are independent, then
	\begin{displaymath}
	\kappa(X \oplus_{q} Y) =
           \kappa(X) \oplus_{q} \kappa(Y) \enspace.
	\end{displaymath}
        \end{enumerate}
        \end{theorem}
	Theorem~\ref{Theorem:CharacterizationOfTsallisEntropy} is a
          direct consequence of
          Theorems~\ref{Theorem:NonextensiveAditivityOfTwoRandomVariables}
          and \ref{Theorem:AxiomsForKN-averages}. 
          This characterization of Tsallis entropy only replaces the
	additivity constraint in the characterization of Shannon
	entropy given by R\'{e}nyi in
	~\cite{Renyi:1960:SomeFundamentalQuestionsOfInformationTheory},
	with pseudo-additivity, which further does not make use
	of the postulate $\kappa(H) + \kappa(-H)=0$. (This postulate is needed to
	distinguish Shannon entropy from R\'{e}nyi entropy). This
	is possible because Tsallis entropy is unique by means of
	KN-averages and under pseudo-additivity.

%         \proof
%         From the Theorem~\ref{Theorem:AxiomsForKN-averages} we have
%         \begin{displaymath}
%         E(H) = {\langle H \rangle}_{\psi} =
%         \psi^{-1}\left(\int \psi \, \ud F \right) \enspace,
%         \end{displaymath}
%         where $\psi$ is strictly monotone and continuous. From the
%         postulate (2) and
%         Theorem~\ref{Theorem:NonextensiveAditivityOfTwoRandomVariables} we
%         have the remaining proof.
%         \endproof

%====================================================================
\section{Conclusions}
\label{Section:Conclusions}

	Passing an information measure through R\'{e}nyi formalism --
	procedure followed by R\'{e}nyi to generalize Shannon entropy
	-- allows one to study the possible generalizations and 
	characterize information measure in the context in terms of
	axioms of quasilinear means. In this paper we studied this
	technique for nonextensive entropy and showed that Tsallis
	entropy is unique under R\'{e}nyi's recipe.
	Considering the attempts to study generalized thermostatistics 
	based on
	KN-averages (for example
	\cite{CzachorNaudts:2002:ThermostatisticsBasedOnKolmogorov-NagumoAverages}),
	the results presented in this paper further the 
	relation between entropic measures and generalized averages.

\section*{References}

\bibliographystyle{unsrt}
\bibliography{papi}

\end{document}